\documentclass[conference]{IEEEtran}
\IEEEoverridecommandlockouts
\usepackage{cite}
\usepackage{amsmath,amssymb,amsfonts}
\usepackage{algorithmic}
\usepackage{graphicx}
\usepackage{textcomp}
\usepackage{xcolor}
\def\BibTeX{{\rm B\kern-.05em{\sc i\kern-.025em b}\kern-.08em
    T\kern-.1667em\lower.7ex\hbox{E}\kern-.125emX}}
\usepackage[T1]{fontenc}    
\begin{document}

\title{Demonstrator Game Showcasing\\Indoor Positioning via BLE Signal Strength
}

\author{
	\IEEEauthorblockN{
		Felix Beierle\IEEEauthorrefmark{1},
		Hai Dinh-Tuan\IEEEauthorrefmark{2},
		and
		Yong Wu\IEEEauthorrefmark{2}
	}
	\IEEEauthorblockA{
		\IEEEauthorrefmark{1}Institute of Clinical Epidemiology and Biometry,
		University of W\"urzburg,	W\"urzburg, Germany\\
		Email: felix.beierle@uni-wuerzburg.de
	}
	\IEEEauthorblockA{\IEEEauthorrefmark{2}Service-centric Networking,
		Technische Universit\"at Berlin,
		Berlin, Germany\\
		Email: hai.dinhtuan@tu-berlin.de, yong.wu1994@gmail.com
	}

}

\maketitle

\begin{abstract}

For a non-technical audience, new concepts
from computer science and engineering
are often hard to grasp.
In order to introduce a general audience
to topics related to Industry 4.0,
we designed and developed a demonstrator game.
The \emph{Who wants to be a millionaire?} style quiz game lets the player experience indoor positioning based on Bluetooth signal strength
firsthand. We found that such an interactive game demonstrator
can function as a conversation-opener
and is useful in helping introduce 
concepts
relevant for many future jobs.

\end{abstract}


\section{Introduction}

Industry 4.0 introduces many concepts from computer science into the factory of the future and will have significant impacts on various aspects of manufacturing, and, in turn, the daily work life of many people.
A lot of people, however, are not familiar with the computer-science-related
mechanisms and concepts that will be implemented.
In order to demonstrate one such concept, indoor positioning, to a general audience,
we design and implement a \emph{Who Wants to be a Millionaire?} style interactive
game that aims at showcasing signal-strength-based indoor positioning
by real-life demonstration.
We utilize a Raspberry Pi as a mobile game screen and four ESP32 microcontrollers that serve as Bluetooth beacons.

\section{Background}

In the smart factory of the future, there will be more and more autonomous mobile robots
on the shop floor. The coordination of those robots can be done by an edge unit in order to optimize the overall workflows \cite{Dinh-TuanMAIAMicroservicesbasedArchitecture2019}.
That coordination can only be feasible when those robots' exact positions are known, leading to the need of a positioning technique.
GPS (Global Positioning System) is well known for precise outdoor positioning, but these signals are likely to be blocked or reflected by walls \cite{peterson1997measuring}, rendering them unusable indoors.
Therefore, there has been a wide array of research on indoor positioning systems. Because of the physical characteristics in indoor environments, different communication technologies such as visible light, ultra-wideband, infrared, etc., have been utilized. However, those implementations still require additional equipment, hence some works focus on using existing infrastructures such as Wi-Fi or Bluetooth signals.

With a similar motivation, in this work, we used Bluetooth Low Energy (BLE) to provide an economical solution for indoor positioning. Bluetooth modules have long been considered as a standard configuration for mobile devices, therefore, although we provide a tablet for the visitor in this specific scenario, this can be easily extended to be an app, which each visitor can download to their smartphone and play the demonstrator game.

\section{Demonstrator Setup}

\subsection{User Perspective}

\begin{figure}[b]
	\center
	\includegraphics[width=0.75\columnwidth]{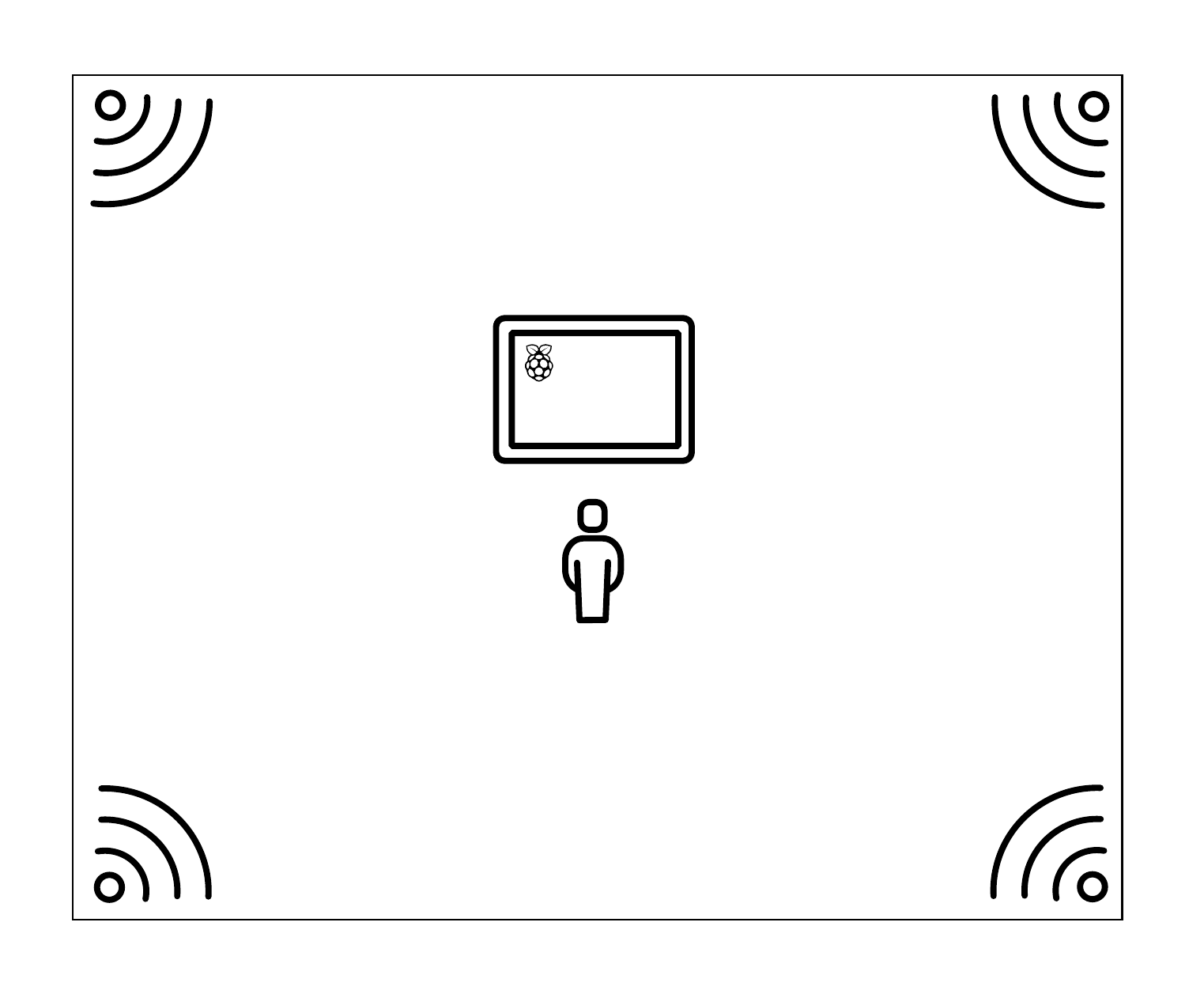}
	\caption{Top-view of the game setup. The player starts in the center of the room, then
		he/she chooses one of four answers to a question shown on the screen
		by walking toward the corresponding corner of the room.}
	\label{fig:setup}
\end{figure}

Figure \ref{fig:setup} shows the demonstrator setup schematically.
The player starts the game in the center of the room, holding
a tablet-like device.
The four corners of the room each are equipped with a Bluetooth beacon.
The game interface displays a set of questions, each with four answer possibilities.
Each answer is associated with a corner of the display as well as the corresponding corner
of the room.
In our demonstrator setup, we additionally color- and number-coded each corner.
In order to select the answer, the user walks toward one of the corners,
which automatically selects the corresponding answer on the screen.
The user presses a button to confirm and sees if the answer he/she chose was correct.

\subsection{Hardware Setup}
\begin{figure}[]
	\center
	\includegraphics[width=0.6\columnwidth]{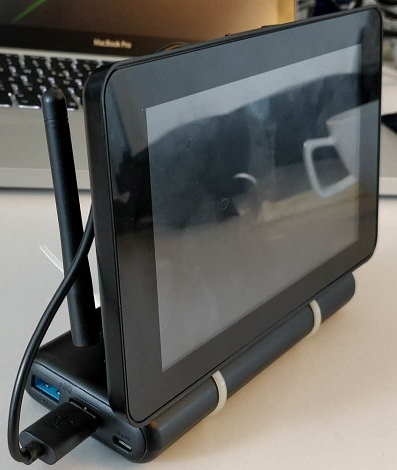}
	\caption{Raspberry Pi 3B with touchscreen, powerbank, and antenna.}
	\label{fig:pi}
\end{figure}

\begin{figure}[]
	\center
	\includegraphics[width=0.6\columnwidth]{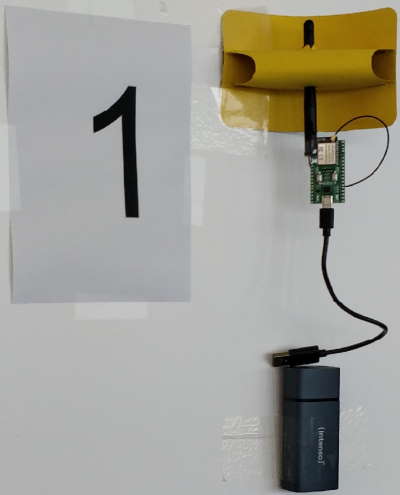}
	\caption{ESP32 microcontroller with antenna, curved signal reflector, and powerbank.}
	\label{fig:beacon}
\end{figure}

Our whole hardware setup including power sources is mobile and can be installed in any room.
There are two types of mobile devices: one Raspberry Pi (Version 3 Model B) and four ESP32 microcontrollers.
We equipped the Raspberry Pi with a 7-inch touchscreen, external antenna, fan, powerbank, and case, effectively making it a type of tablet computer (see Figure \ref{fig:pi}).
We equipped each of the four ESP32 microcontrollers with small, portable powerbanks,
external antennas and self-made, colored reflectors for directing the Bluetooth signal. 
Figure \ref{fig:beacon} shows the final installation of a beacon in one of the corners of our demonstrator room.


\subsection{Software Setup}
\begin{figure}[h]
	\center
	\includegraphics[width=0.955\columnwidth]{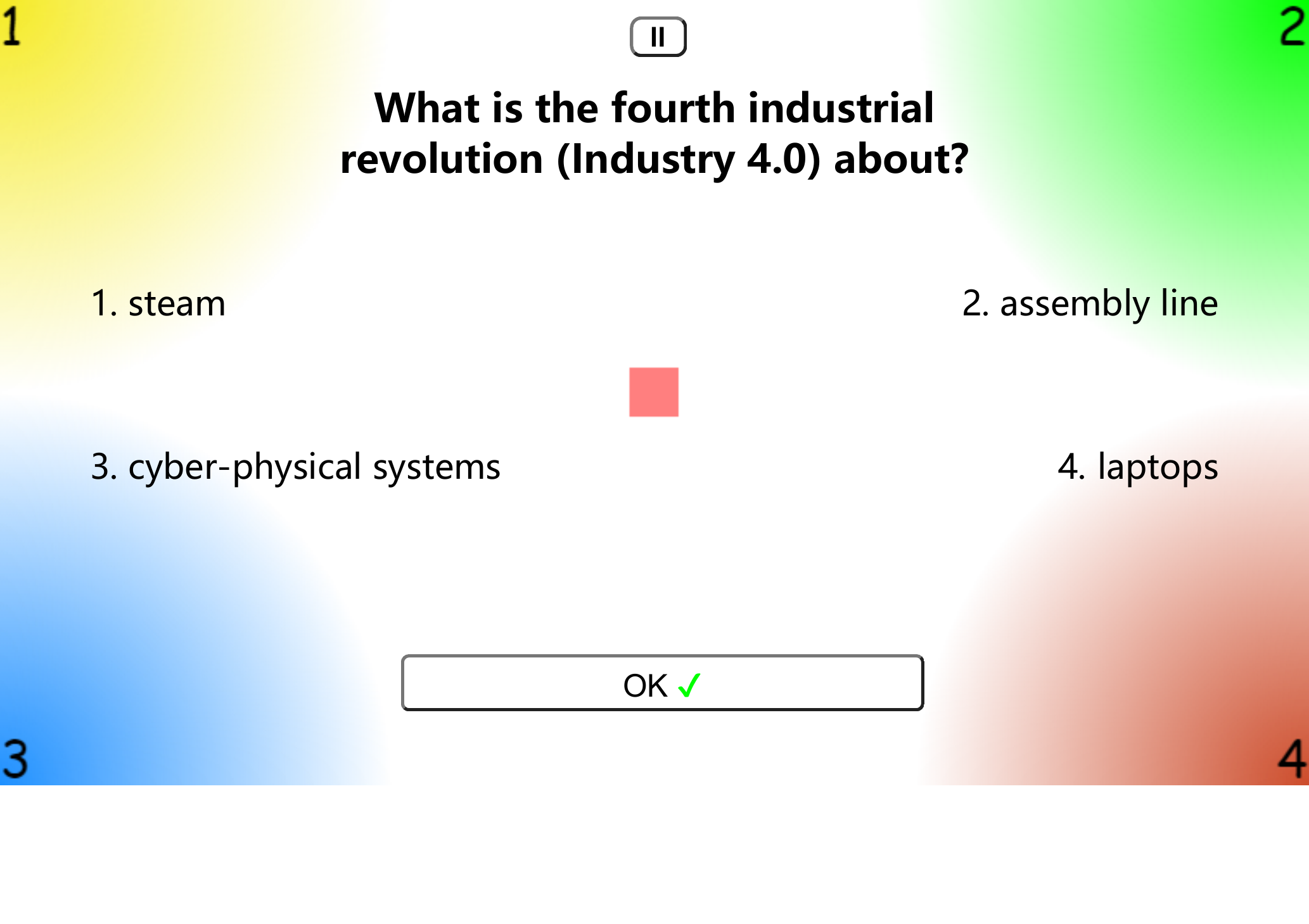}
	\caption{Example screenshot of the game. The red square in the center indicates the position of the player as determined by the Bluetooth signal strength of the four beacons in the four corners of the room.}
	\label{fig:screenshot}
\end{figure}

We developed and deployed a small program on the ESP32s to turn them into Bluetooth beacons, broadcasting pre-defined Universally Unique Identifiers (UUIDs).
The code for the Raspberry Pi consists of two parts: Python code for
scanning Bluetooth signals and JavaScript/HTML code for the game frontend.
Based on the received signal strength, we apply a heuristic to estimate the distance to each beacon.
In order to reduce the effect of anomalous values, we use a moving average filter
of the last ten received broadcasts.
We defined a threshold for being close enough to a beacon, which then selects the chosen
answer in the frontend.
On the screen, besides questions and possible answers, there is an icon indicating the position
of the player in relation to the four corners of the room.
Figure \ref{fig:screenshot} shows a screenshot of the game with an example question.

\section{Impressions/Discussion}

We implemented and tested the game at a public science event in Berlin,
Germany on June 15, 2019. 
Many people from various backgrounds and with various ages played the game,
and overall, the feedback was very positive.
We found that the such an interactive game serves as a good demonstrator 
to explain about signal strength, indoor positioning, Industry 4.0, and
related concepts to a general audience.

Less than a year after our presentation, during the coronavirus pandemic,
Bluetooth signal strength was also used in another context.
Here, some contact-tracing apps use the Bluetooth signal between
smartphones for storing information on which devices have been in proximity,
providing the possibility to warn users  when they have been in contact with
people who tested positive later.

\section*{Acknowledgment}
We are grateful for the support provided by Axel K\"upper.

\bibliographystyle{IEEEtran}

\begin{thebibliography}{1}
\providecommand{\url}[1]{#1}
\csname url@samestyle\endcsname
\providecommand{\newblock}{\relax}
\providecommand{\bibinfo}[2]{#2}
\providecommand{\BIBentrySTDinterwordspacing}{\spaceskip=0pt\relax}
\providecommand{\BIBentryALTinterwordstretchfactor}{4}
\providecommand{\BIBentryALTinterwordspacing}{\spaceskip=\fontdimen2\font plus
\BIBentryALTinterwordstretchfactor\fontdimen3\font minus
  \fontdimen4\font\relax}
\providecommand{\BIBforeignlanguage}[2]{{%
\expandafter\ifx\csname l@#1\endcsname\relax
\typeout{** WARNING: IEEEtran.bst: No hyphenation pattern has been}%
\typeout{** loaded for the language `#1'. Using the pattern for}%
\typeout{** the default language instead.}%
\else
\language=\csname l@#1\endcsname
\fi
#2}}
\providecommand{\BIBdecl}{\relax}
\BIBdecl

\bibitem{Dinh-TuanMAIAMicroservicesbasedArchitecture2019}
H.~{Dinh-Tuan}, F.~Beierle, and S.~Rodriguez~Garzon, ``{{MAIA}}: {{A
  Microservices}}-based {{Architecture}} for {{Industrial Data Analytics}},''
  in \emph{Proc. 2019 {{IEEE International Conference}} on {{Industrial Cyber
  Physical Systems}} ({{ICPS}})}.\hskip 1em plus 0.5em minus 0.4em\relax
  {IEEE}, May 2019, pp. 23--30.

\bibitem{peterson1997measuring}
B.~Peterson, D.~Bruckner, and S.~Heye, ``Measuring gps signals indoors,'' in
  \emph{Proceedings of the 10th International Technical Meeting of the
  Satellite Division of The Institute of Navigation (ION GPS 1997)}, 1997, pp.
  615--624.

\end{thebibliography}

\end{document}